\documentclass[reprint,aps,prl,twocolumn,superscriptaddress,nobibnotes,floatfix]{revtex4-1}%

\usepackage{amsmath}
\usepackage{amssymb}
\usepackage{natbib}
\usepackage{graphicx}
\usepackage{xcolor}
\usepackage[unicode=true, breaklinks=false, pdfborder={0 0 1}, backref=false, colorlinks=true, linkcolor=blue, urlcolor=blue, citecolor=blue]{hyperref}

\setcitestyle{numbers,square}

\begin{document}

\title{Optical Manipulation of Domains in Chiral Topological Superconductors}

\author{Tao Yu}
\affiliation{Max Planck Institute for the Structure and Dynamics of Matter,
	Luruper Chaussee 149, 22761 Hamburg, Germany}
	\author{Martin Claassen}
\affiliation{Department of Physics, University of Pennsylvania, Philadelphia, PA 19104, USA}
\affiliation{Center for Computational Quantum Physics, Simons Foundation Flatiron Institute, New York, NY, USA}
\author{Dante M. Kennes}
\affiliation{
Institut f\"ur Theorie der Statistischen Physik, RWTH Aachen, 52056 Aachen, Germany}
\affiliation{Max Planck Institute for the Structure and Dynamics of Matter,
	Luruper Chaussee 149, 22761 Hamburg, Germany}
\author{Michael A. Sentef}
\affiliation{Max Planck Institute for the Structure and Dynamics of Matter,
	Luruper Chaussee 149, 22761 Hamburg, Germany}

\date{\today}

\begin{abstract}
Optical control of chirality in chiral superconductors bears potential for future topological quantum computing applications. When a chiral domain is written and erased by a laser spot, the Majorana modes around the domain can be manipulated on ultrafast time scales. Here we study topological superconductors with two chiral order parameters coupled via light fields by a time-dependent real-space Ginzburg-Landau approach. Continuous optical driving, or the application of supercurrent, hybridizes the two chiral order parameters, allowing one to induce and control the superconducting state beyond what is possible in equilibrium. We show that superconductivity can even be enhanced if the mutual coupling between two order parameters is sufficiently strong. Furthermore, we demonstrate that short optical pulses with spot size larger than a critical one can overcome a counteracting diffusion effect and write, erase, or move chiral domains. Surprisingly, these domains are found to be stable, which might enable optically programmable quantum computers in the future.
\end{abstract}
\maketitle

\textit{Introduction}.---Chiral superconductors spontaneously break time-reversal symmetry and host topologically protected chiral Majorana edge modes \cite{Read_Green,Majorana_RMP}, thus bearing potential for applications in quantum computing \cite{quantum_computing,lian2018}. 
The superconductor's chirality itself can be a useful degree of freedom, much like spin \cite{spintronics1,spintronics2}, valley \cite{valleytronics1,valleytronics2}, and other quantum states \cite{superconducting_qubit}, which may allow for quantum information processing \cite{Martin_NP}. It is therefore desirable to develop means of controlling chiral superconductors and their associated Majorana edge modes, preferably on ultrafast time scales. 
Dynamical symmetry breaking by optical driving \cite{Floquet1,Floquet2,Martin_NP}, or supercurrents \cite{SHG1,SHG2}, has been suggested as an efficient way to achieve this goal. 

It was recently predicted that optical switching of chirality in a bulk chiral superconductor can be achieved by a joint effect of homogeneous linearly and circularly polarized optical pulses \cite{Martin_NP}. The local and ultrafast manipulation of Majorana modes, however, requires the creation and annihilation of chiral domains by a laser spot of finite size. It is not a priori obvious that this can be achieved, since locally perturbed order parameters can diffusively relax back to their original state in a system with multi-component superconducting instabilities. Here we microscopically derive the time-dependent Ginzburg-Landau (TDGL) equations \cite{Kopnin,Gorkov} in real space from a prototype model on a honeycomb lattice \cite{RMP,graphene_NP,Honerkamp,RMT1,RMT2}, which could be realized, as suggested by recent experiments, in highly doped graphene \cite{high_doping_exp_1,high_doping_exp_2}, or  twisted bilayer graphene and other van der Waals materials \cite{TBL1,TBLFRG1,TBBN,TBL3,TBL4,TBWeS,TBLFRG2,TBDBG,TBL_Xu}. Further works on chiral superconducting materials, which will follow a similar phenomenology as discussed here, include experimental evidence for chiral $p\pm ip$-wave superconductivity in materials, such as UPt$_3$ \cite{UPt1,UPt2,UPt3} and UTe$_2$ \cite{UTe}. The perhaps most notable candidate material has been Sr$_2$RuO$_4$ with proposals of triplet $p$-wave pairing \cite{SrRuO1,SrRuO2}, but a more recent investigation ruled this out \cite{SrRuO3_challenge}, and there is now evidence for $d$- or potentially even $g$-wave instabilities \cite{sr2ruo4-new1,sr2ruo4-new2,kivelson}. Importantly, our TDGL theory phenomenologically but transparently describes the dynamics of coupled order parameters of general chiral superconductors in both spatially homogeneous and inhomogeneous scenarios. 
 
 \begin{figure}[t]
 	\begin{center}
 		{\includegraphics[width=8.25cm,trim=0.5cm 0cm 0.5cm 0cm]{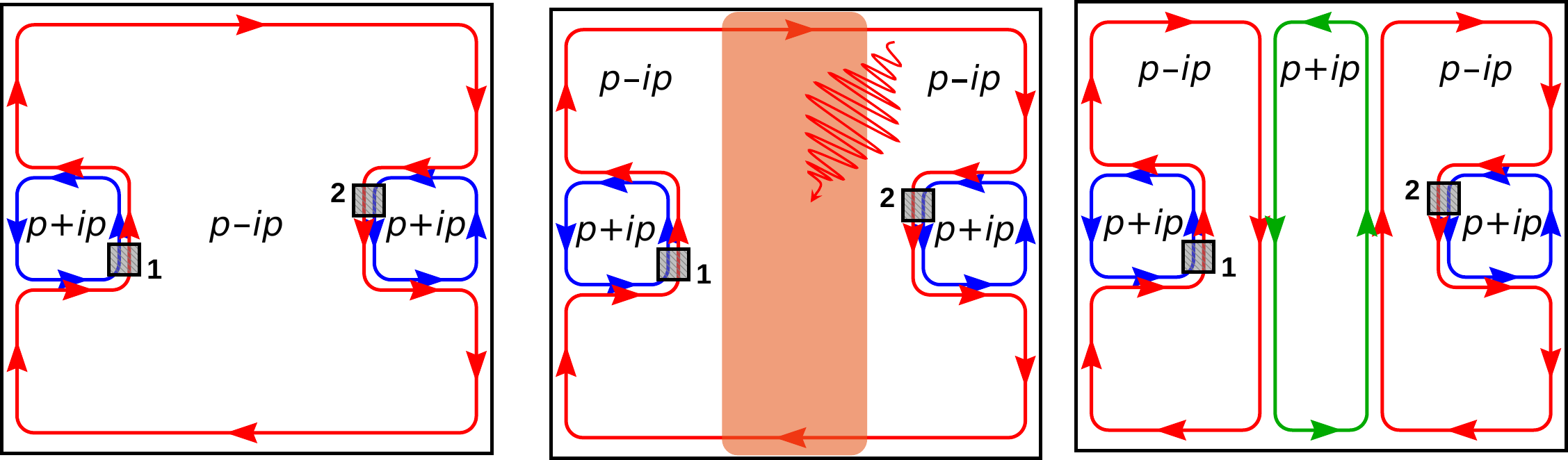}}
 		\caption{Braiding of Majorana fermions using optically-engineered domains in a $p+ip$ superconductor. A single electron is injected at the two leads (gray shaded boxes) into the domain walls between $p+ip$ and $p-ip$ states, which carry two co-propagating Majorana modes. Depending on the topography of domains, the fractionalization and propagation of these electrons along the Majorana boundary modes implements either a Hadamard gate (left) or an identity operation (right) on the charging states of the leads. Using optical manipulation, one can switch between these two gates on ultrafast time scales (center).}   
 		\label{fig:braiding}
 	\end{center}
 \end{figure}

Our main results are two-fold. On the one hand, we find that because the two chiral order parameters are coupled via the optical vector potential as a consequence of angular-momentum conservation, a {\it continuous driving} (or a static supercurrent) can hybridize the two order parameters. Provided that the coupling between the two order parameters is sufficiently strong, superconductivity can even be enhanced beyond their equilibrium values. On the other hand, we demonstrate that a {\it short pulse} of linearly polarized light can drive the two order parameters to be close in magnitude. Due to the accompanying amplitude-mode oscillation (Higgs mode) \cite{Nambu,Higgs_s1,Higgs_s2,Higgs_d,Higgs_cea,Tsuji,Kemper,Yu_gauge}, the balance between the two chiral order parameters depends on the precise time at which the pulse is switched off with respect to this oscillation. The probability for a switch from one order to the other to occur is roughly $1/2$. A second circularly polarized pulse can be used to push the switching to favor one order parameter over the other and allows for a reliable switching \cite{Martin_NP}.  
Finally, for the spatially inhomogeneous situation, we predict that there is a minimal spot size required to overcome the diffusion effect, which counteracts the switching. Above a critical spot size switching can be obtained in a stable fashion, allowing for the optical creation or annihilation of chiral domains of multi-component superconductors \cite{RMP,Martin_NP}, such as the widely studied chiral $p\pm ip$ superconductivity. A combination of creation and annihilation can be used to move the domain.  

Since the induced chiral domain is stable after application of optical pulses, manipulation of Majorana modes at the boundary of such domains is possible. This could be used to optically program quantum logic gates.
Figure.~\ref{fig:braiding} depicts a possible implementation of a Hadamard gate \cite{Hadamard_1,Hadamard_2,Hadamard_3} using the Majorana modes of optically-engineered domains in a $p+ip$ superconductor. Analogous to the mechanism described in Ref.~\cite{lian2018}, the single-electron charging states of two leads can be used to encode quantum bits of information \cite{Martin_NP}. Upon injection of a single electron at a $p+ip$/$p-ip$ domain wall boundary that carries a pair of Majorana modes or equivalently a single chiral complex fermion mode, the electron can propagate along the edge channel. Crucially, the superconducting \textit{sample} boundary hosts only a single Majorana mode, necessitating a fractionalization of the electron into Majorana fermions. By optically-induced choice of domain topography, these propagating Majorana modes can then implement braiding operations, realizing a Hadamard gate on charging states of leads \cite{lian2018}.

\textit{Model and formalism.}---To be specific, we consider a chiral $d\pm id$ superconductor on a honeycomb lattice. The general results, however, will carry over to other chiral superconductors (with two-component order parameters), such as $p\pm ip$ superconductors. The tight-binding $t$-$J$ model Hamiltonian reads \cite{RMP,graphene_NP,Honerkamp,RMT1,RMT2} 
	\begin{align}
	\nonumber
	\hat{H}_{t\mbox{-}J}&=-t\sum_{\langle i,j\rangle,\sigma}(\hat{a}^{\dagger}_{i\sigma}\hat{b}_{j\sigma}+\hat{b}^{\dagger}_{j\sigma}\hat{a}_{i\sigma})-\mu\sum_{i,\sigma}(\hat{a}_{i\sigma}^{\dagger}\hat{a}_{i\sigma}+\hat{b}_{i\sigma}^{\dagger}\hat{b}_{i\sigma})\\
	&-J\sum_{\langle i,j\rangle}\hat{h}_{ij}^{\dagger}\hat{h}_{ij},
	\label{Hamiltonian}
	\end{align}
	with $\hat{a}_{\sigma}$ and $\hat{b}_{\sigma}$ being the annihilation operators of electrons with spin $\sigma=\{\uparrow,\downarrow\}$ on the A and B sublattices. The hopping amplitude between nearest-neighbor lattice sites is denoted by $t$. The system is assumed to be doped near the van Hove singularity with the 
	chemical potential $\mu\sim t$  \cite{graphene_NP,Honerkamp,TBL1,TBLFRG1,TBBN,TBL3,TBL4,high_doping_exp_1,high_doping_exp_2,TBL_Xu}. We consider a nearest-neighbor spin exchange interaction $J$ between spin singlets with creation operators $\hat{h}^{\dagger}_{ij}=\left(\hat{a}^{\dagger}_{i\uparrow}\hat{b}^{\dagger}_{j\downarrow}-\hat{a}^{\dagger}_{i\downarrow}\hat{b}^{\dagger}_{j\uparrow}\right)$ with sites $i\in \;$A and $j\in \;$B \cite{graphene_NP,Honerkamp}. Based on this Hamiltonian, we derive the Ginzburg-Landau (GL) Lagrangian for the superconducting order parameters \cite{path_integral} (see Supplemental Material for details \cite{supplement}).  
	
	With the order parameters  $\eta_1({\bf r},t)$, $\eta_2({\bf r},t)$ of the $(d_{x^2-y^2}+id_{xy})$-, $(d_{x^2-y^2}-id_{xy})$-superconducting tendency, respectively, 
	the GL Lagrangian density reads 
		\begin{align}
		\nonumber
		{\cal L}_{\rm eff}({\bf r})&=\sum_{\mu=1,2}\Gamma_{\mu}\eta_{\mu}^*({\bf r})D_t\eta_{\mu}({\bf r})+\sum_{\mu}\Lambda_{\mu}\left|D_t\eta_{\mu}({\bf r})\right|^2\\
		&+a\sum_{\mu}|\eta_{\mu}({\bf r})|^2+b\sum_{\alpha=x,y}\sum_{\mu}\left|D_{\alpha}\eta_{\mu}({\bf r})\right|^2\nonumber\\
		\nonumber
		&+e_1(1-i\sqrt{3})\big[D_+\eta_2^*({\bf r})\big]\big[D_+ \eta_1({\bf r})\big]\\
		\nonumber
		&+e_1(1+i\sqrt{3})\big[D_+^*\eta_2({\bf r})\big]\big[D_+^* \eta^*_1({\bf r})\big]\nonumber\\
		&+f_1\left(|\eta_1|^2+|\eta_2|^2\right)^2+f_2\left(|\eta_1|^2-|\eta_2|^2\right)^2.
		\label{GL_Lagrangian}
		\end{align}
		The $s$-wave order parameter is disregarded here, as it is not energetically favored \cite{supplement,Martin_NP,graphene_NP,Honerkamp} and does not affect the main conclusions presented here (its inclusion is analyzed in Supplemental Material \cite{supplement}). Here, $D_t\equiv \partial_t-({2e}/{i\hbar})\varphi({\bf r},t)$,  $D_{\alpha}=\partial_{\alpha}-({2e}/{i\hbar c}){\bf A}_{\alpha}({\bf r},t)$ and $D_{\pm}=D_{x}\pm iD_y$ are the covariant derivatives that respect the gauge invariance of the Lagrangian \cite{RMP}, 
	where $\varphi({\bf r},t)$ and ${\bf A}({\bf r},t)$ are the scalar and vector potentials of the electromagnetic field, and $c$ is the speed of light. The first $\Gamma$-term is of the Gross-Pitaevskii type, and is responsible for dissipation \cite{Kopnin,Gorkov} back to equilibrium. The second $\Lambda$-term is of the Klein-Gordon type, which leads to collective-mode dynamics, such as the amplitude or Higgs mode, that is defined by the fluctuation of the order-parameter amplitude \cite{Nambu,Tsuji,Yu_gauge}. Here the parameters $\Gamma_{\mu}$ and $\Lambda_{\mu}$ are treated phenomenologically and are allowed to be different for the two $d\pm id$ order parameters only if time-reversal symmetry is explicitly broken, for instance by a circularly polarized laser field, following Ref.~\cite{Martin_NP}. The other coefficients $\{a<0,b,e_1,f_1>0,f_2=-f_1/3\}$ are all real numbers that are microscopically calculated through the Hamiltonian~(\ref{Hamiltonian}) \cite{supplement}. Estimated by graphene's material parameters $t=2.7$~eV, the bonding length $|{\bf b}|=2.46/\sqrt{3}$~\AA \cite{graphene_RMP}, $\mu=0.9t$ and $J=0.25t$ \cite{Honerkamp,Martin_NP,supplement}, $a=-10^{-3}/|{\bf b}|^2~{\rm meV}^{-1}\cdot{\rm m}^{-2}$, $b=800~{\rm meV}^{-1}$, $e_1=150~{\rm meV}^{-1}$,  $f_1=2.1/|{\bf b}|^2~{\rm meV}^{-3}\cdot{\rm m}^{-2}$ at $T=1.5$~K. These parameters correspond to a critical temperature $T_c=4.7$~K \cite{supplement}. We choose $\Gamma=5|a|\hbar/(k_BT_c)$, with a realistic nanosecond relaxation time for small order-parameter fluctuations, and $\Lambda=-a\hbar^2/(2|\eta_0|^2)$, with the superconducting gap $\eta_0=\sqrt{-a/(2f_1+2f_2)}=0.02~{\rm m eV}$  \cite{supplement}, following Ref.~\cite{Gorkov}. However, importantly, this GL Lagrangian can also be constructed based entirely on symmetry considerations \cite{RMP}, independent of microscopic details, rendering our resutls applicable to a wide range of physical situations.
	
	When $f_2<0$, the two chiral 
	($d\pm id$)-waves are degenerate superconducting ground states \cite{graphene_NP,Honerkamp}, which is the case for Hamiltonian~(\ref{Hamiltonian}) (see Supplemental Material  \cite{supplement}). Otherwise, $d_{x^2-y^2}=(\eta_1+i\eta_2)/2$ and $d_{xy}=(\eta_1-i\eta_2)/2$ waves are the ground states.
	As the two order parameters $\eta_1$ and $\eta_2$ are coupled in the $e_1$-terms, obeying angular-momentum conservation, 
	 their interplay can be optically controlled. From $\delta L/\delta \eta_{1,2}^*({\bf r})=0$ \cite{Kopnin} we obtain the TDGL equations 
\begin{align}
\nonumber
&\left(\begin{matrix}
\Gamma_1&0\\
0&\Gamma_2
\end{matrix}\right) D_t\left(\begin{matrix}
\eta_1\\
\eta_2
\end{matrix}\right)+\left(\begin{matrix}
\Lambda_1&0\\
0&\Lambda_2
\end{matrix}\right) D_t^*D_t\left(\begin{matrix}
  \eta_1\\
   \eta_2
\end{matrix}\right)\\
\nonumber
&+\left(\begin{matrix}
a-b\sum_{\alpha=x,y}D_{\alpha}^2&-e_1(1+i\sqrt{3})\tilde{D}_{-}D_{-}\\
-e_1(1-i\sqrt{3})\tilde{D}_{+}D_{+}&a-b\sum_{\alpha=x,y}D_{\alpha}^2 
\end{matrix}\right)\left(\begin{matrix}
\eta_1\\
\eta_2
\end{matrix}\right)\\
&+\left(\begin{matrix}
{\mathcal F}_+(|\eta_1|^2,|\eta_2|^2)&0\\
0&{\mathcal F}_-(|\eta_1|^2,|\eta_2|^2)
\end{matrix}\right)\left(\begin{matrix}
\eta_1\\
\eta_2
\end{matrix}\right)=0,
\label{TDGL}
\end{align}
where $\tilde{D}_{\nu}=\partial_{\nu}+({2e}/{i\hbar c}){\bf A}_{\nu}({\bf r})$, $\tilde{D}_{\pm}=\tilde{D}_x\pm i\tilde{D}_y$ and ${\cal F}_{\pm}(|\eta_1|^2,|\eta_2|^2)\equiv 2f_1(|\eta_1|^2+|\eta_2|^2)\pm 2f_2(|\eta_1|^2-|\eta_2|^2)$. We choose a gauge with $\varphi=0$ and consider the dynamics under vector potentials ${\bf A}_{x,y}({\bf r},t)$ that are in general spatially and temporally dependent, which we shall specify in the following. 

\textit{Mechanism of chirality switching.}--- We first discuss the coupling between the two chiral order parameters through light in the homogeneous case. Here, we can analyze the chirality switching mechanism analytically. The coupling of the two order parameters is $e_1({\bf A}_x^2-{\bf A}_y^2)\rightarrow 0$ for the circularly polarized laser within the rotating-wave approximation. Therefore, as also confirmed by our numerical calculation, circularly polarized light by itself cannot cause chirality switching. We therefore focus here on the linearly polarized laser first.

\begin{figure}[t]
	\begin{center}
		{\includegraphics[width=4.43cm]{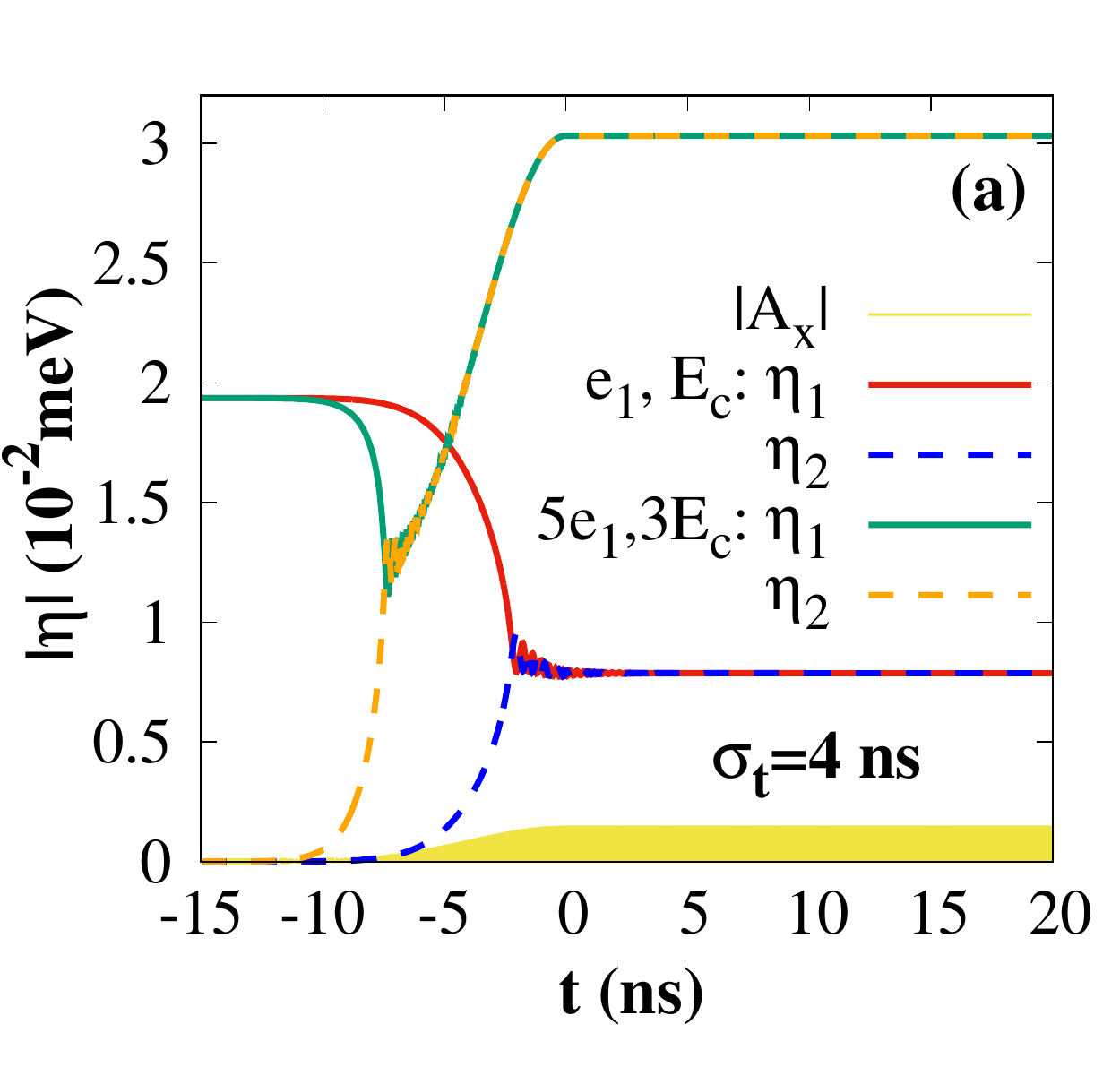}}
		\hspace{-0.42cm} 
		{\includegraphics[width=4.43cm]{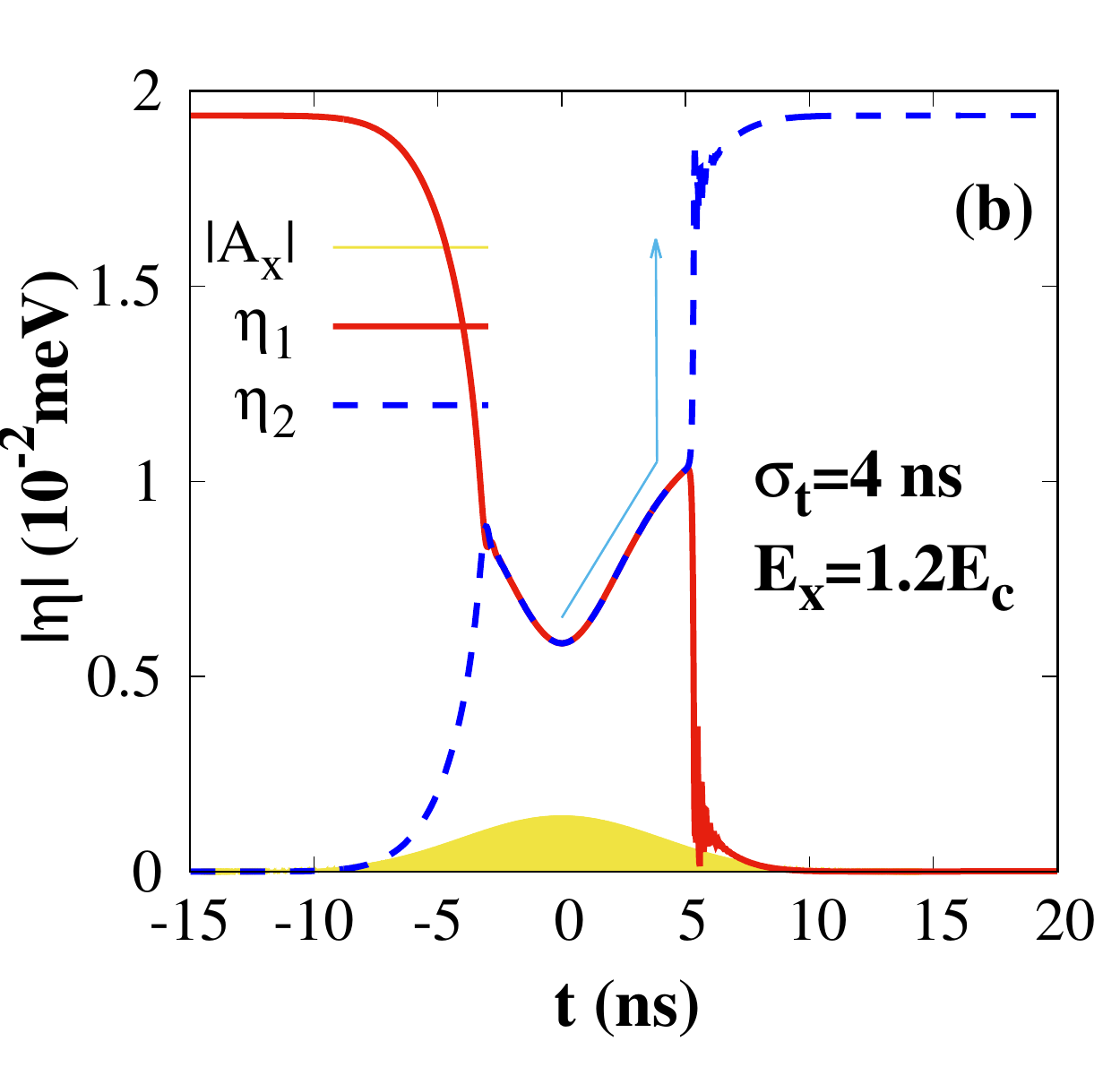}}
		\hspace{-0.3cm}{\includegraphics[width=4.43cm]{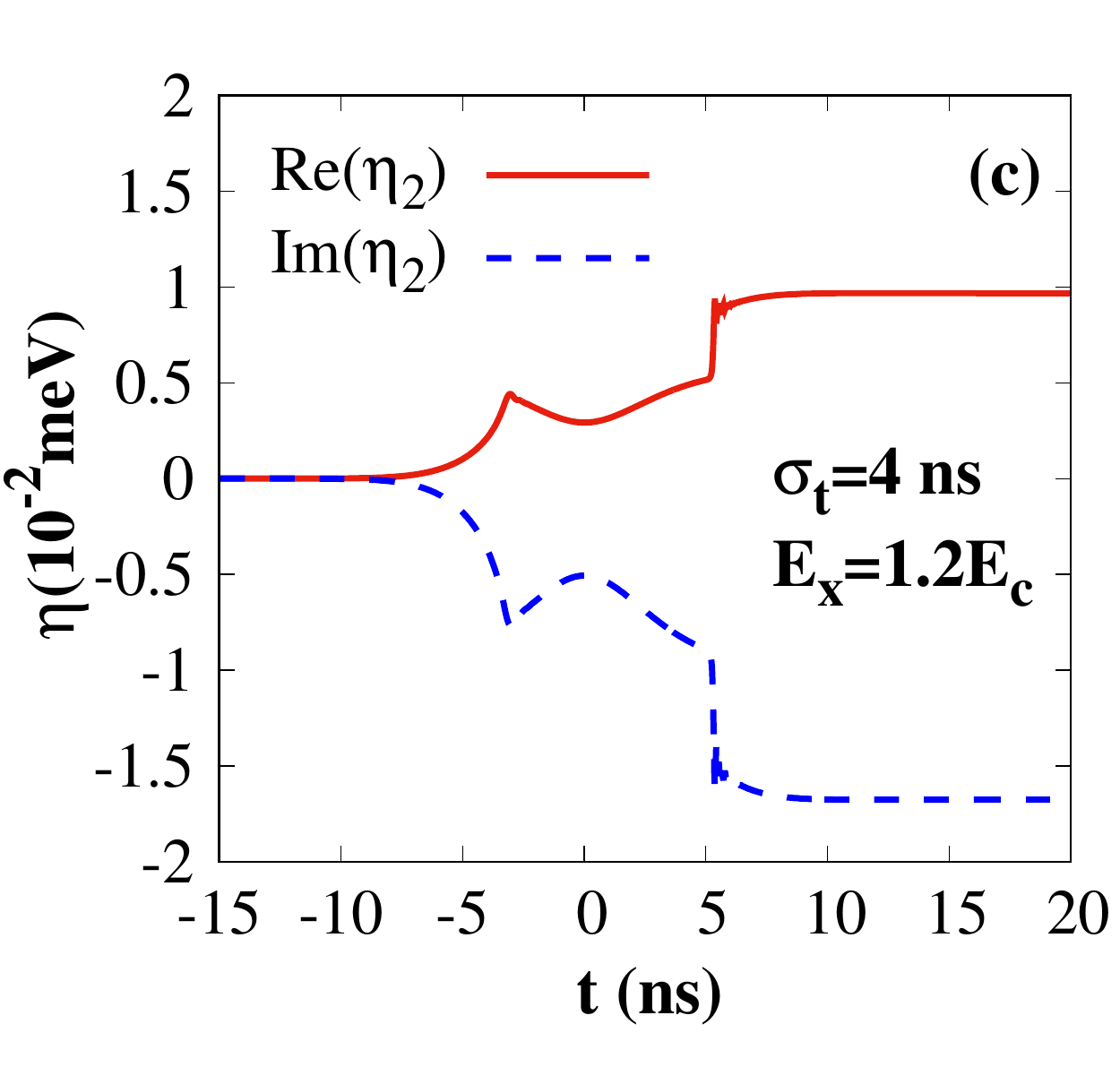}}
		\hspace{-0.41cm}
		{\includegraphics[width=4.43cm]{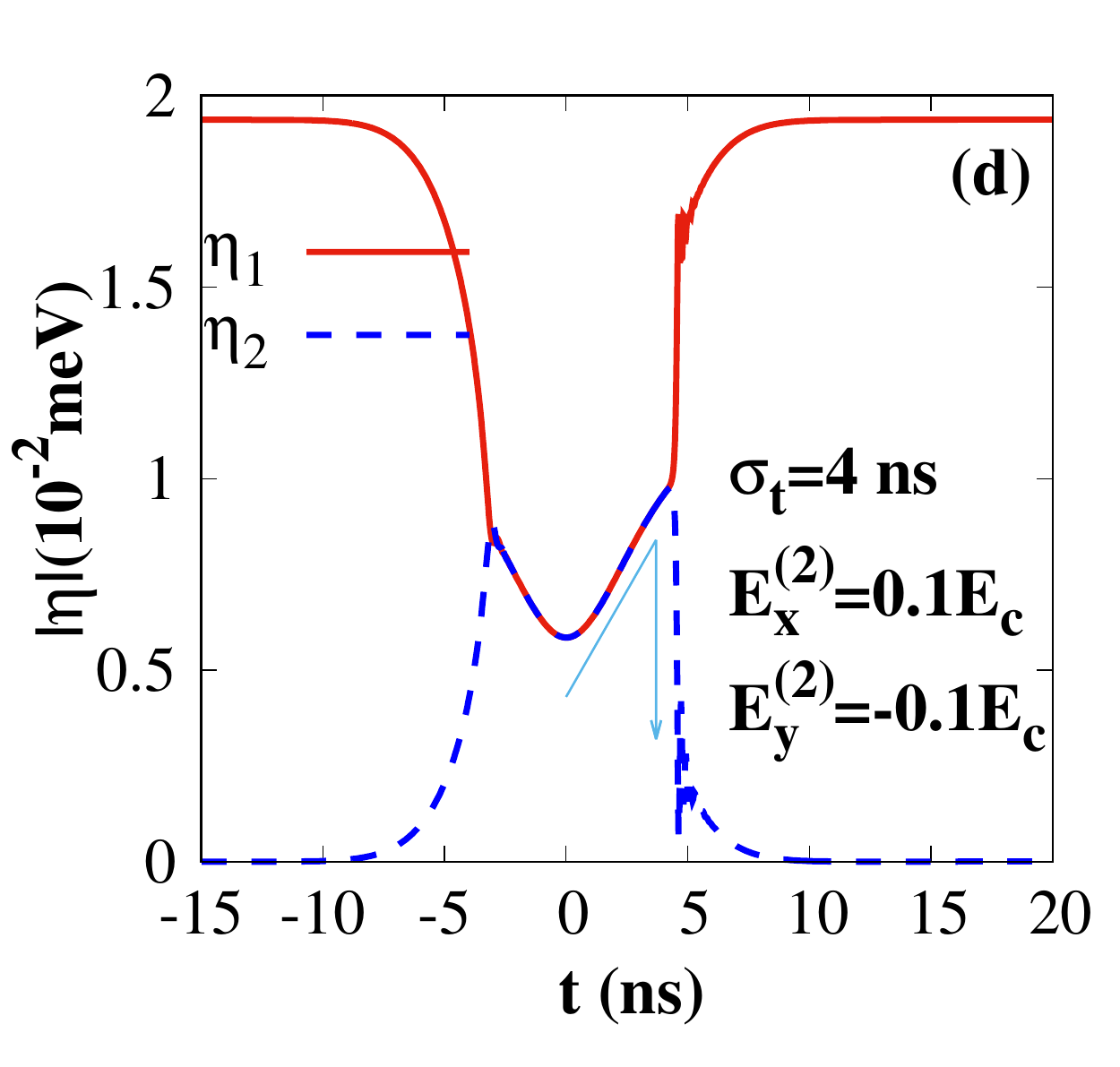}}
		\caption{Optical engineering of order parameters by homogeneous optical fields. (a) Hybridization of two chiral order parameters by continuous linearly polarized optical fields. An arbitrary value $5e_1$ for the order-parameter coupling is used for illustration of superconductivity enhancement. Optical switching of chirality is shown in (b) by order parameter magnitude and (c) by real and imaginary components of $\eta_2$. In (d), a second left-circularly polarized laser with time delay $\sigma_t$ relative to the first linearly polarized laser is applied to reverse the chirality, while the right-circularly polarized laser can retain the chirality (not shown). Parameters used for calculation are given in the figure and text.}
		\label{fig:dynamics}
	\end{center}
\end{figure}

We first analyze the hybridization of order parameters by a {\it continuous} optical driving of frequency $\omega$ and electric-field amplitude $\mathbf{E}$, say ${\bf A}_x=-({c}/{\omega}){\bf E}_x\cos(\omega t)$ (equivalent to sourcing  a constant supercurrent). When the two chiral order parameters are driven to the same magnitude, we use the ansatz $\tilde{\eta}_2=\tilde{\eta}_1e^{i\phi}$ to solve the TDGL equation (\ref{TDGL}) and find $\phi=-\pi/3$ and magnitude 
\begin{equation}
|\tilde{\eta}_{1,2}|=\sqrt{\frac{1}{4f_1}\left(-a+2(2e_1-b)\left(\frac{e E_x}{\hbar \omega}\right)^2 \right)},
\label{hybridization}
\end{equation}
which is tunable by the field strength. This implies that by driving superconductivity is suppressed when $2e_1<b$, otherwise it is enhanced 
due to the strong spatial fluctuation between the two order parameters.
Such a hybridization is confirmed in Fig.~\ref{fig:dynamics}(a) with adiabatically turned on continuous driving of frequency $\hbar \omega=20$~meV with
${\bf A}_x=-({c}/{\omega}){\bf E}_x\cos(\omega t)\exp\left({-t^2/(2\sigma_t^2)}\right)$ for duration time $\sigma_t=4$~ns for $t<0$ and $-({c}/{\omega}){\bf E}_x\cos(\omega t)$ for $t>0$.
This hybridization also implies the existence of a critical supercurrent that induces nodes in the superconducting gap before superconductivity is destroyed, providing a unique feature of chiral superconductors. 	

 We now turn to the discussion of chirality switching with {\it short pulses} in the homogeneous case. The similar magnitude of order parameters through driving is essential for the chirality switching, which happens when the passive order parameter, initially absent, overcomes the initially active one. The order parameters can be driven to be close only when the field strength is larger than a critical value $E_c\approx (\hbar\omega/e)\sqrt{-a/(b-2e_1)}=1~{\rm kV/cm}$, estimated by setting $|\tilde{\eta}_{1,2}|\rightarrow \eta_0/2$ in Eq.~(\ref{hybridization}).
 Figure~\ref{fig:dynamics}(b) confirms that, when the two chiral order parameters are driven to be close by a linearly polarized laser with $E_x=1.2E_c$, switching of chirality happens after the pulse, accompanied by an oscillation of the order parameters. This indicates that the Higgs mode activation causes a temporal fluctuation that provides an opportunity for the passive order parameter to overcome its inertia. The final state $\eta_2$ after the switching has a phase shift $-\pi/3$ relative to the initial one $\eta_1$ [Fig.~\ref{fig:dynamics}(c)], as expected from order-parameter hybridization [Eq.~(\ref{hybridization})]. Nevertheless, when we further increase the field strength or change the laser pulse duration, the switching does not always happen. Half of the time the order parameter simply relaxes back to the initial state with no switching, which we demonstrate in thousands of calculated cases with different switch off parameters.
	
An explicit breaking of time-reversal symmetry by a second circularly polarized laser, with time delay $\sigma_t$ to the first one, turns out to be key to obtain full control of the switching process \cite{Martin_NP}. Although the circularly polarized light cannot couple the two order parameters directly, as discussed above, it may influence their damping since single-particle excitations or the environment could be polarized by the laser \cite{Floquet_engineering_1,Floquet_engineering_2}. Neither of these effects are directly included in the TDGL framework, but require a more microscopic treatment, such as Bogoliubov-de-Gennes equations \cite{Martin_NP}. We therefore assume that the damping rate depends on the chirality of the laser by a minimal phenomenological ansatz
$\Gamma_{1}\rightarrow \Gamma_1(1+\delta_c E^{(2)}_xE^{(2)}_y/E_c^2)$ and $\Gamma_{2}\rightarrow \Gamma_2(1-\delta_c E^{(2)}_xE^{(2)}_y/E_c^2)$, 
where a tiny dimensionless $\delta_c\sim 0.1>0$ is used.
We indeed realize complete control of the switching by the chirality of the second laser pulse in all the regimes with this ansatz, recovering the discovery of Ref.~\cite{Martin_NP}, as shown in Fig.~\ref{fig:dynamics}(d) for weak fields $E^{(2)}_{x}=-E^{(2)}_{y}\sim 0.1E_c$ of left-handed circularly polarized light.

\begin{figure}[t]
	\begin{center}
		{\includegraphics[width=8.6cm]{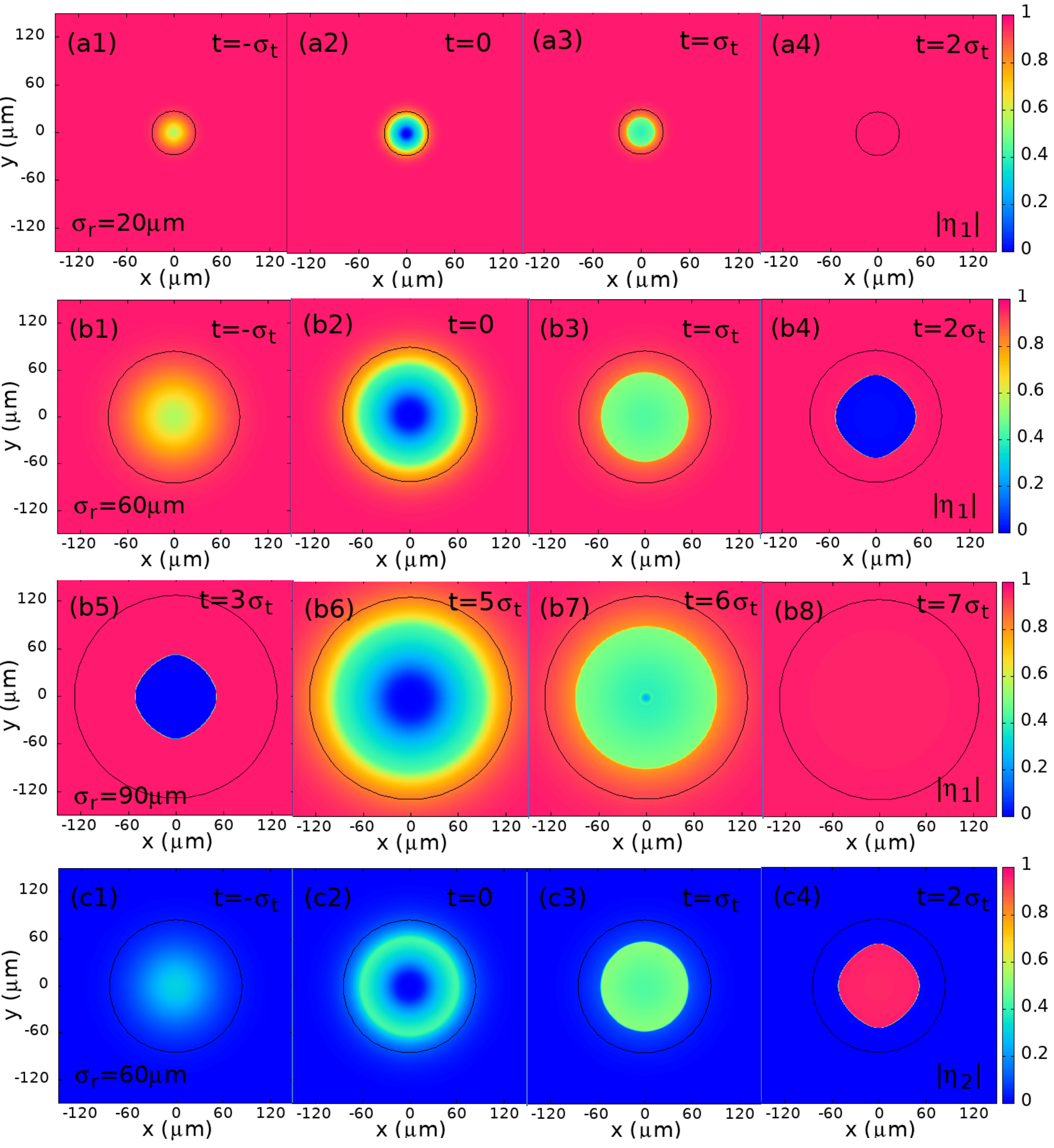}}
		\caption{Optical engineering of order parameters (normalized by $\eta_0$) in real space by laser spots. The spot size $\sqrt{2}\sigma_r$ is indicated by black circles. (a1)-(a4) plot the snapshots of $|\eta_1|$ under a laser of small size with $\sigma_r=20~\mu$m. The order parameter relaxes to its equilibrium after the pulse [(a4)], not switched. (b1-b4) and (c1-c4) plot the evolutions of $|\eta_1|$ and $|\eta_2|$, respectively, when the spot size is larger, with $\sigma_r=60~\mu$m. Switching is achieved [(b4),(c4)] and stable afterwards. (b5-b6) illustrate the optical erasure of a chiral domain in real space by linearly polarized laser at $t\in [4\sigma_t,6\sigma_t]$, with spot size $\sigma_r=90~\mu$m, and a second pulse of left-circular polarization at $t\in [5\sigma_t,7\sigma_t]$, which favors the switching from $\eta_2$ to $\eta_1$. Parameters for calculation are given in the text.}
		\label{fig:real_space}
	\end{center}
\end{figure}

\textit{Optical engineering of chiral domain}.---We now address finite-spot--sized laser pulses, in order to investigate the possibility of writing, erasing, or moving chiral domains for implementations in potential future quantum computers (Fig.~\ref{fig:braiding}, see Supplemental Material for implementing Hadamard gates using Majorana edge modes \cite{supplement}). We now restrict the laser field to a spot of size $\sqrt{2}\sigma_r$ by using fields ${\bf A}_y(t)=-({c}/{\omega}){\bf E}_y\sin(\omega t)  \exp[-{t^2}/({2\sigma_t^2)}]
\exp[-{r^2}/({2\sigma_r^2})]$. With a right-circularly polarized laser $E^{(2)}_x=E^{(2)}_y=0.1E_c$ of $\sigma_t=2$~ns delay to this linearly polarized field $E_y=1.5E_c$, we compute the superconductor dynamics with the TDGL equations in real space. We illustrate the results with two representative spot sizes (indicated by the black circles) $\sqrt{2}\sigma_r=20\sqrt{2}$ and $60\sqrt{2}$~$\mu$m in Fig.~\ref{fig:real_space}. These spot sizes are much larger than the coherence length of the order parameters $\xi\sim\sqrt{-b/(2a)}\approx 0.1~{\mu}$m \cite{Landau,supplement}. When the spot size is small ($\sigma_r=20~\mu$m), switching is not achieved after the pulse, as shown in Fig.~\ref{fig:real_space}(a1)-(a4). The switching becomes possible, however, when the spot size is larger ($\sigma_r=60~\mu$m), as shown in Fig.~\ref{fig:real_space}(b4) and (c4).

In fact, as long as the spot size $\sigma_r\gtrsim 20~\mu$m, we always find the chirality switching with the domain size smaller than the spot size, indicating the existence of a critical size $\sigma_r^{(c)}$ of photo-induced chiral domain. The switching is always reversed when the handedness of second pulse is reversed. From the TDGL equations, an order-parameter fluctuation relaxes to equilibrium on a time scale $\tau \sim 3$~ns (compare Fig.~\ref{fig:dynamics}(b)); on the other hand, the excited spatial fluctuation propagates with the Higgs-mode propagation speed $v=\sqrt{\Lambda/b}\sim 5.3$~km/s. We may thereby estimate the critical spot size as $\sigma_r^{(c)}\sim v\tau=16~{\mu}$m, agreeing reasonably with the numerical calculation. It is important to note that this critical spot size is two orders of magnitude larger than the superconducting coherence length in our model.

The ability of optical erasure of a chiral domain is essential for application, since the combination of creation and annihilation can move a chiral domain on demand, potentially allowing for the braiding of Majorana modes \cite{Hadamard_1,Hadamard_2,Hadamard_3,quantum_computing,Majorana_RMP}. On the basis of the chiral domain in Fig.~\ref{fig:real_space}(b4) and (c4), we now apply similar optical pulses but using a left-circularly polarized laser that favors the switching from $\eta_2$ to $\eta_1$. The diffusion of order parameters at the edge leaves a ring of chiral domain with a small size (see Supplemental Material \cite{supplement}), which, however, can be erased entirely by pulses of larger size, such as $\sigma_r=90~\mu$m, shown in Fig.~\ref{fig:real_space}(b5-b8).  

Again, in real space the phase of $\eta_2$ has a shift $-\pi/3$ with respect to $\eta_1$, and hence the order parameter at the center of the domain wall is neither $d_{x^2-y^2}$ nor $d_{xy}$, but a superposition. The free energy is increased, as the domain wall costs additional energy \cite{supplement}, indicating that the chiral domain may be energetically metastable. However, the chiral order parameters carry opposite angular momenta and topological winding numbers such that their direct conversion breaks angular momentum conservation, seemingly not possible in the absence of magnetic fields \cite{stability1,stability2}.    
Indeed, the photo-induced chiral domain appears to be stable since it does not vanish after the pulse in the TDGL computation, robust to order-parameter fluctuation. 

\textit{Discussion.}---Phenomenological GL approaches solely rely on system symmetry and hence do not strongly depend on the underlying mechanism for chiral superconductivity \cite{RMP}. The parameters therein depend on the environment such as the substrate and phonons, which may play a role in the superconducting mechanism \cite{Wu_phonon,screening} and, as bosonic and fermionic baths, can influence the switching processes. On the one hand they can relax hot quasiparticles to suppress heating effects. On the other hand, they can enhance the damping rate $\Gamma$ \cite{Martin_NP,Kopnin} and therefore tend to decelerate the chirality switching and favor a larger critical size of the laser spots for local switching.

We have demonstrated local control of chirality by focused laser pulses that can write, erase and move chiral domains, and addressed the way to realize optically programmable quantum logic gates. For realistic unconventional superconductors, the required field strengths are within reach: a field of one to several kV/cm of tens of terahertz is large enough to cause sufficient coupling between order parameters and trigger the switching, such that heating effects might be small. The required minimal size of the laser spot is tens of micrometers to overcome the diffusion when writing a chiral domain. We find that an enhancement of superconductivity may be possible in particular materials with sufficiently strong coupling of order parameters by optical driving. The hybridization of chiral order parameters is measurable by a scanning tunneling microscope, which can track the nodes in the gap that are created in this case \cite{STM_RMP}. 
The local chirality of the superconductor can be measured via Kerr rotation with spatial resolution \cite{Kerr_rotation_1,Kerr_rotation_2}, or via the anomalous Hall effect \cite{AHE_single,AHE_multi,AHE_Yu,AHE_PRX}. Our study also suggests that unique features in the optical response of chiral superconductors may be useful for identifying the superconducting state of magic-angle twisted bilayer graphene and other van der Waals materials \cite{TBL1,TBLFRG1,TBBN,TBL3,TBL4,TBWeS,TBLFRG2,TBDBG,TBL_Xu}.


\vskip0.25cm 
\begin{acknowledgments}
TY thanks Damian Hofmann for help on high-performance computing. TY and MAS acknowledge financial support by Deutsche Forschungsgemeinschaft through the Emmy Noether program (SE 2558/2-1). The Flatiron Institute is a Division of the Simons Foundation. DMK acknowledges support by the Deutsche
Forschungsgemeinschaft (DFG, German Research Foundation) via RTG 1995 and Germany’s Excellence Strategy - Cluster of Excellence Matter and Light for Quantum Computing (ML4Q) EXC 2004/1 - 390534769. We acknowledge support from the Max Planck-New York City Center for Non-Equilibrium Quantum Phenomena. 
\end{acknowledgments}

\end{document}